\documentclass[10pt]{iopart}
\usepackage{iopams}

\usepackage{color}
\usepackage{graphicx}
\usepackage{sidecap} 
\usepackage[square,sort&compress,numbers]{natbib}
\usepackage{url}
\usepackage{xspace}
\usepackage{hyperref} 

\newcommand{\aligo}{Advanced LIGO\gdef\aligo{aLIGO\xspace}\xspace}

\newcommand{\gw}[1][]{gravitational wave#1 (GW#1)\renewcommand{\gw}[1][]{GW##1\xspace}\xspace}
\providecommand{\ns}[1][]{neutron star#1 (NS#1)\renewcommand{\ns}[1][]{NS##1\xspace}\xspace}
\renewcommand{\ns}[1][]{neutron star#1 (NS#1)\renewcommand{\ns}[1][]{NS##1\xspace}\xspace}
\newcommand{\cw}[1][]{continuous wave#1 (CW#1)\renewcommand{\cw}[1][]{CW##1\xspace}\xspace}
\newcommand{\gpu}[1][]{graphics processing unit#1 (GPU#1)\renewcommand{\gpu}[1][]{GPU##1\xspace}\xspace}
\newcommand{\Fstat}[1][]{$\F$-statistic#1\xspace}
\newcommand{\sft}[1][]{Short Fourier Transform#1 (SFT#1)\renewcommand{\sft}[1][]{SFT##1\xspace}\xspace}
\newcommand{\snr}[1][]{signal-to-noise ratio#1 (SNR#1)\renewcommand{\snr}[1][]{SNR##1\xspace}\xspace}

\newcommand{\sw}[1]{\texttt{#1}\xspace}
\newcommand{\lalsuite}{\sw{LALSuite}}
\newcommand{\lalpulsar}{\sw{LALPulsar}}
\newcommand{\pycuda}{\sw{pyCUDA}}
\newcommand{\pyfstat}{\sw{PyFstat}}
\newcommand{\numpy}{\sw{NumPy}}

\newcommand{\pgm}{Prix:2011qv}
\newcommand{\citepgm}{\cite{\pgm}\xspace}
\newcommand{\pyfref}{Ashton:2018ure,ashton_gregory_2018_1243931}
\newcommand{\citepyfstat}{\cite{\pyfref}\xspace}

\newcommand{\eto}{\ensuremath{\mathrm{e}^}}     
\newcommand{\definedas}{:=}

\newcommand{\F}{\mathcal{F}}
\newcommand{\Fmn}{\F_{mn}}
\newcommand{\M}{\mathcal{M}}

\newcommand{\Hyp}{\mathcal{H}}
\newcommand{\Gauss}{\mathrm{\MakeUppercase{G}}}
\newcommand{\Signal}{{\mathrm{\MakeUppercase{S}}}}
\newcommand{\Transient}{{\mathrm{t}}}
\newcommand{\Transsig}{{\mathrm{\Transient\Signal}}}

\newcommand{\HypG}{\Hyp_\Gauss}
\newcommand{\HypTS}{\Hyp_{\Transsig}}

\newcommand{\Amp}{\mathcal{A}}
\newcommand{\Dop}{\lambda}
\newcommand{\TP}{\mathcal{T}}

\newcommand{\freq}{f}
\newcommand{\fdot}{{\dot{\freq}}}
\newcommand{\fddot}{\ddot{\freq}}

\newcommand{\win}{\varpi}
\newcommand{\winexp}{\win_\mathrm{exp}}
\newcommand{\winrect}{\win_\mathrm{rect}}

\newcommand{\pInfo}{\mathcal{I}}
\newcommand{\prob}[2]{P\left(#1\middle|#2,\pInfo\right)}

\newcommand{\Ad}{\widehat{A}}
\newcommand{\Bd}{\widehat{B}}
\newcommand{\Cd}{\widehat{C}}
\newcommand{\Dd}{\widehat{D}}

\newcommand{\SFT}{{\mathrm{SFT}}}
\newcommand{\Tsft}{T_\SFT}
\newcommand{\Tdata}{T_{\mathrm{data}}}
\newcommand{\Tobs}{T_\mathrm{obs}}
\newcommand{\Nsft}{N_\SFT}

\newcommand{\Nto}{N_{t_0}}
\newcommand{\Ntau}{N_{\tau}}
\newcommand{\res}{d}
\newcommand{\resto}{\res{t_0}}
\newcommand{\restau}{\res{\tau}}

\newcommand{\tomin}{t_{0\min}}
\newcommand{\tomax}{t_{0\max}}
\newcommand{\taumin}{\tau_{\min}}
\newcommand{\taumax}{\tau_{\max}}

\newcommand{\Fmap}{{\F\mathrm{map}}}
\newcommand{\cost}{\mathfrak{c}}
\newcommand{\costFmaprect}{\cost^\mathrm{r}_\Fmap}
\newcommand{\costFmapexp}{\cost^\mathrm{e}_\Fmap}

\newcommand{\costrectPGM}{\mathfrak{c}_\mathrm{r}}
\newcommand{\costexpPGM}{\mathfrak{c}_\mathrm{e}}
\newcommand{\costrectsums}{\cost^\mathrm{r}_\mathrm{sums}}
\newcommand{\costexpsums}{\cost^\mathrm{e}_\mathrm{sums}}
\newcommand{\costF}{\mathfrak{c}_\F}
\newcommand{\costmarg}{\mathfrak{c}_\mathrm{marg}}
\newcommand{\Nsums}{N_\mathrm{sums}}
\newcommand{\Nsumsrect}{\Nsums^\mathrm{r}}
\newcommand{\Nsumsexp}{\Nsums^\mathrm{e}}
\newcommand{\costSFTs}{\mathfrak{c}_\mathrm{SFTs}}

\newcommand{\fgw}{\freq_{\mathrm{GW}}}
\newcommand{\fspin}{\freq_{\mathrm{spin}}}

\newcommand{\dcc}{LIGO-P1800031-v6\xspace}
\newcommand{\glasgow}{University of Glasgow, School of Physics and Astronomy, Kelvin Building, Glasgow G12 8QQ, Scotland, United Kingdom}
\newcommand{\aei}{Max Planck Institut für Gravitationsphysik (Albert Einstein Institut), 30161 Hannover, Germany}
\newcommand{\monash}{Monash Centre for Astrophysics, School of Physics and Astronomy, Monash University, VIC 3800, Australia}

\begin{document}

\title[GPU transient \Fstat]{Faster search for long gravitational-wave transients: GPU implementation of the transient \Fstat}

\author{David Keitel$^1$ and Gregory Ashton$^{2,3}$}
\address{$^1$\glasgow}
\address{$^2$ \aei}
\address{$^3$ \monash}
\ead{david.keitel@ligo.org}

\vspace{10pt}
\begin{indented}
 \item[]\dcc [draft version: 15 June 2018]
\end{indented}

\begin{abstract}
The \Fstat is an established method
to search for continuous gravitational waves
from spinning neutron stars.
Prix et al. \citep[(2011)]{\pgm}
introduced a variant
for transient quasi-monochromatic signals.
Possible astrophysical scenarios for such transients
include glitching pulsars,
newborn neutron stars
and accreting systems.
Here we present a new implementation of the transient \Fstat,
using \pycuda to leverage the power
of modern graphics processing units (GPUs).
The obtained speedup
allows efficient searches over much wider parameter spaces,
especially when using more realistic transient signal models
including time-varying
(e.g. exponentially decaying)
amplitudes.
Hence, it can enable comprehensive coverage
of glitches in known nearby pulsars,
improve the follow-up of outliers from continuous-wave searches,
and might be an important ingredient
for future blind all-sky searches for unknown neutron stars.
\end{abstract}

%
%
%
%
%

\section{Introduction}

Spinning \ns[s],
when non-axisymmetrically deformed,
emit weak
but potentially detectable
\gw[s]~\cite{Prix:2009oha}.
Many searches~\cite{Riles:2017evm}
with the LIGO and Virgo detectors~\cite{TheLIGOScientific:2014jea,TheVirgo:2014hva} 
focus on \cw signals that are persistent over a whole observation run,
but there are also scenarios for shorter signals from transiently perturbed \ns[s].
If those signals are slowly evolving in frequency
and last on the time scale of hours to months,
analysis methods adapted from \cw searches are well suited to their detection.
In~\citepgm
(hereafter also referred to as `PGM'),
the astrophysical motivation for such transient signals was discussed
and a matched-filter search method proposed.
It is based on the established \Fstat,
which was introduced in~\cite{Jaranowski:1998qm,Cutler:2005hc}
and used in many \cw searches~\citep[recently e.g. in][]{Aasi:2014ksa,Zhu:2016ghk,Abbott:2017pqa}.

Matched-filter searches for weak signals from unknown sources
(or those with imperfectly known parameters)
are computationally very expensive
since a wide parameter space needs to be densely covered with templates.
Starting from a typical \cw search that covers a certain parameter space in
signal frequency, spindown and sky location
but assumes a constant signal amplitude,
the addition of new unknown parameters to describe the transient evolution
further increases computational cost.

However, the attractiveness of the transient \Fstat algorithm from~\citepgm
is that it starts from time-discretised quantities already computed
for the standard \cw \Fstat
and then only needs to take partial sums of these to study
the set of possible transient signals.
Still, for long total observation times
the evaluation of these partial sums can easily dominate over
the original computational cost,
especially if the templates have a non-trivial amplitude evolution,
e.g. exponential decay.

The task of multiple partial sums of some input data
can obviously benefit from massive parallelisation.
Here we present a straightforward translation of the algorithm from~\citepgm
to \pycuda code~\cite{Kloeckner:2012_pyc} running on \gpu[s].
It is implemented in the framework of the \pyfstat package~\citepyfstat.\footnote{Latest
source code and examples also available from \url{https://gitlab.aei.uni-hannover.de/GregAshton/PyFstat/}.}

In the following, we briefly review the formalism from~\citepgm
to define the transient \Fstat (\autoref{sec:form}),
then describe its \pycuda implementation (\autoref{sec:impl}).
We test the speed and memory requirements (\autoref{sec:tests})
and compare with the original CPU implementation
from \lalsuite~\cite{lalsuite}.
The paper ends with a brief discussion (\autoref{sec:concl})
of how the achieved speedup
widens the scope of feasible searches
for long \cw-like gravitational wave transients.
This includes enabling a comprehensive
coverage of glitch events in nearby known pulsars,
improving the sensitivity of all-sky \cw searches
through following up more outliers with transient analyses,
and the potential use as an ingredient
in future blind all-sky searches
for unknown disturbed \ns[s].

\section{Formalism}
\label{sec:form}

We present a straightforward \pycuda implementation
of PGM's `atoms-based' transient \Fstat algorithm
from Appendix A1 of~\citepgm.
It is based on a discretised method to compute the overall \Fstat
introduced in~\cite{Williams:1999nt}
and described in detail in \cite{Prix:cfsv2}.

The \Fstat
(for transient or continuous signals)
is essentially a likelihood-ratio test
for a time series $x(t)$,
comparing a signal hypothesis
\begin{equation}
 \label{eq:hypTS}
  \HypTS :  x(t) = n(t) + h(t,\Dop,\Amp,\TP)
\end{equation}
against the alternative hypothesis of pure Gaussian noise,
\begin{equation}
 \label{eq:hypG}
 \HypG :  x(t) = n(t) \,.
\end{equation}
The waveform model
\mbox{$h(t,\Dop,\Amp,\TP) = \win(t,\TP)\,h(t,\Dop,\Amp)$}
for a slowly-evolving signal
separates into a transient window function $\win(t,\TP)$
and the standard \cw waveform $h(t,\Dop,\Amp)$.
The latter depends on
a set of phase evolution parameters
\mbox{$\Dop=\{\alpha,\delta,\freq,\fdot,\fddot,\dots\}$}
(sky position,
frequency,
and frequency derivatives or `spindowns')
and on four amplitude parameters
\mbox{$\Amp=\{h_0,\cos\iota,\psi,\phi_0\}$}.
See~\cite{\pgm,Jaranowski:1998qm,Prix:cfsv2} for details
on these parameters.
For the transient part,
we currently consider
either rectangular or exponential
window functions,
with parameters \mbox{$\TP=\{t_0,\tau\}$}
where
$t_0$ is the start time of a signal
and $\tau$ is a duration parameter:
\begin{equation}
  \label{eq:winrect}
  \winrect(t,t_0,\tau) \definedas \left\{\begin{array}{ll}
      1 & \mbox{if } t \in [t_0,t_0+\tau]\\
      0 & \mbox{otherwise}\,,
      \end{array}\right.
\end{equation}
\begin{equation}
  \label{eq:winexp}
  \winexp(t,t_0,\tau) \definedas \left\{\begin{array}{ll}
      \eto{-(t - t_0)/\tau} & \mbox{if } t \in [t_0,t_0+3\tau]\\
      0 & \mbox{otherwise}\,.
      \end{array}\right.
\end{equation}
The cutoff of $\winexp$ at $3\tau$ was introduced in~\citepgm
in the understanding that the \snr after this point will be negligible.

The (transient) \Fstat is then proportional to the log odds
between $\HypTS$ and $\HypG$,
after maximising over $\Amp$
(or marginalising, see~\cite{\pgm,Prix:2009tq,Keitel:2013wga} for details):
\begin{equation}
 \label{eq:F_OSG}
 \eto{\F(x,\Dop,\TP)} \propto \frac{\prob{\HypTS}{x,\Dop,\TP}}{\prob{\HypG}{x}} \,.
\end{equation}
It can be written as
\begin{equation}
 \label{eq:F_xMx}
 \F(x,\Dop,\TP) = \frac{1}{2} x'_\mu(\Dop,\TP) \, \M'^{\mu\nu}(\Dop,\TP) \, x'_\nu(\Dop,\TP) \,,
\end{equation}
where the indices $\mu$, $\nu$ run over the four amplitude parameters $\Amp$,
$\M'^{\mu\nu}$ is the antenna pattern matrix,
$x'_\mu$ are projections of the data onto the model waveforms,
and the prime denotes transient windowing.
(See Eqs. (32--36) of~\citepgm.)

The standard algorithm used in \Fstat searches for continuous signals
splits a data set starting at $T_0$ and of length $\Tobs$
into several \sft[s] of length $\Tsft$.
\cite{Prix:cfsv2} describes how to approximate \eref{eq:F_xMx}
from the per-SFT, per-detector discretised versions
of $\M'^{\mu\nu}$ and $x'_\mu$;
in practice we consider the equivalent set of quantities
\mbox{$\{a_j,b_j,F_{aj},F_{bj}\}$}
as the \emph{atoms} of our \Fstat computation,
where the $j$ index runs over SFTs.

The $a_j$ and $b_j$ atoms are summed up to yield
the discretised antenna pattern matrix elements
$\Ad$, $\Bd$, $\Cd$
\citep[defined in Eq. (130) of][]{Prix:cfsv2}
and
their determinant
\mbox{$\Dd = \Ad\,\Bd - \Cd^2$},
and together with the summed data-dependent
complex quantities $F_a$, $F_b$
\citep[Eq.(129) of][]{Prix:cfsv2}
they yield the \Fstat as:
\begin{eqnarray}
 \label{eq:F_FaFb}
 \F(x,\Dop,\TP)  = \Dd^{-1} \, ( &   \Bd \, [ \, \Re^2(F_a) + \Im^2(F_a) \, ]
                        +  \Ad \, [ \, \Re^2(F_b) + \Im^2(F_b) \, ] \\
                      &- 2\Cd \, [ \, \Re(F_a)\Re(F_b) + \Im(F_a)\Im(F_b) \, ] \,
                     ) \,. \nonumber
\end{eqnarray}
(All quantities on the right hand side
are understood
as depending on $\lambda$ and $\TP$, too.)

For persistent \cw[s],
this is evaluated summing all atoms over the full $\Tobs$.
To search for transient signals,
we define a grid in $\{t_0,\tau\}$ space
indexed by $m$ for the $t_0$ dimension and $n$ for the $\tau$ dimension.
We indicate the resolutions of this grid as $\resto$ and $\restau$;
a natural choice is \mbox{$\resto=\restau=\Tsft$}
though a coarser or even variable sampling is also possible.
Then our goal is to compute,
for each $\Dop$ and a specific window choice $\win$,
the matrix
\begin{equation}
 \label{eq:Fmn}
 \Fmn(\Dop) \definedas \F(x,\Dop,\win,t_{0\,m},\tau_n) \,, 
\end{equation}
which we also refer to as the
\emph{transient $\F$-statistic map}.
Computing it this way is convenient
because the set of atoms
\mbox{$\{a_j,b_j,F_{aj},F_{bj}\}$}
is only computed once,
over the full $\Tobs$,
and this is already done for the \cw \Fstat anyway.
Subsequently, the transient $\Fmn$ map is obtained
by evaluating \eref{eq:F_FaFb}
for partial sums of the atoms.

\section{Implementation}
\label{sec:impl}

\pyfstat is a python package
primarily developed for the MCMC follow-up~\citepyfstat
of \cw candidates,
but it also provides
general modular access to \cw search functionality
in the \lalpulsar package
(written in C)
of the \lalsuite~\cite{lalsuite} collection,
called through SWIG C-to-python wrappers.
For the transient \Fstat,
we first call a standard algorithm
for computing the \cw \Fstat over the whole data set~\cite{Prix:cfsv2}\footnote{As of
the writing of this paper, documentation is available at: \\
\url{https://lscsoft.docs.ligo.org/lalsuite/lalpulsar/group___compute_fstat__h.html}},
which takes care of the data read-in,
barycentring,
and computation of the per-SFT matched-filter atoms.
The only change is that we ask the \sw{ComputeFstat()} routine
to also return the atoms.

The input data for computing the transient \Fstat map $\Fmn$
consists then of only the atoms
(a set of vectors of $\Nsft$ elements each)
and the parameters describing a transient window function
and grid in $\{t_0,\tau\}$ space.
The inputs are the 3 real vectors $a^2(t)$, $b^2(t)$ and $a(t) \cdot b(t)$
and the 2 complex vectors $F_a(t)$, $F_b(t)$.
These are transferred to the \gpu
as a $7\times\Nsft$ real matrix.

The basic idea of massively-parallelised computation
on a \gpu is to run a grid of identical kernels,
each processing the subset of data
identified by the kernel's \mbox{(multi-)index}.
We provide two structurally different kernels
for rectangular and exponential windows.
To account for the general case
where resolutions in $t_0$ or $\tau$ different from $\Tsft$
might be desirable,
or where there are gaps in the data,
we use $\Nto$ and $\Ntau$ for the number of grid points
in each dimension,
which need not be equal to each other nor to $\Nsft$.

In the \textbf{rectangular case},
an obvious optimisation was already pointed out in \citepgm
and is implemented in \lalpulsar:
For each starting time $t_{0\,m}$,
one can compute $\Fmn$ for all durations $\tau_{n}$
by keeping the partial sums of each atom up to each $\tau_{n'}$
in memory
and only adding the atoms with index $n'+1$ in the next step.
It would thus be wasteful to run a full
$\Nto \times \Ntau$ grid of kernels on the \gpu,
and instead we only launch $\Nto$ kernels,
each of which internally loops over $\tau$
and keeps the partial sums in local memory.

In the \textbf{exponential case},
no such simple trick is possible,
since the contribution to each partial sum
at each timestep
includes amplitude-weight factors (see Eq.~\eref{eq:winexp})
depending on the $\tau$ currently being evaluated.
Hence, we employ a brute-force grid of 
$\Nto \times \Ntau$ kernels on the \gpu,
each of which only computes the partial sums for a single $\Fmn$.

In both cases,
the last steps,
still done inside the \gpu kernel,
are to compute the antenna pattern matrix determinant $\Dd$
and the transient $\Fmn$-statistic from Eq.~\eref{eq:F_FaFb}.

\section{Tests}
\label{sec:tests}

In this section,
we describe tests of the speedup
obtained with the \pycuda version,
its memory requirements,
and its numerical faithfulness to the original implementation.

\subsection{Speed}
\label{sec:timing}

We have tested the speed of the \pycuda implementation relative
to the standard \lalpulsar code
on several systems.
These all have Intel CPUs:
a laptop with a Core i5-6200U at 2.30\,GHz,
a workstation with a Xeon X5675 at 3.07\,GHz
and two LIGO Caltech cluster nodes with
Xeons E5-2630 and E5-2650 at 2.20\,GHz each.
The \pycuda code was benchmarked on 
several \gpu[s] from the Nvidia GeForce GTX family
(1050, 1060, 1070 and 1080Ti,
with 2--11\,GB RAM)
and on a Nvidia Tesla V100-PCIE (16\,GB RAM),
all installed on the same workstation and cluster nodes.

We consider observation times $\Tobs$
from 1 hour up to 1 year,
with no gaps in the data.
Gaussian noise and a transient signal with
\mbox{$\tau=0.5\,\Tobs$}
are simulated through \pyfstat,
though the speed of calculating \Fstat[s]
does not depend on whether the data contains a signal.
The \sft[s] are taken at
\mbox{$\Tsft=1800$\,s}
and $\Fmn$ is sampled at
\mbox{$\resto=\restau=\Tsft$}
over a grid of
\mbox{$t_0  \in [T_0, \Tobs - 2\Tsft]$}
and
\mbox{$\tau \in [2\Tsft, \Tobs]$}.
The upper limit on $t_0$ and lower limit on $\tau$ are set
because the low-level implementation
requires at least 2 \sft[s] per $\Fmn$ computation.

\begin{figure}
 \begin{center}
  \includegraphics[width=\columnwidth]{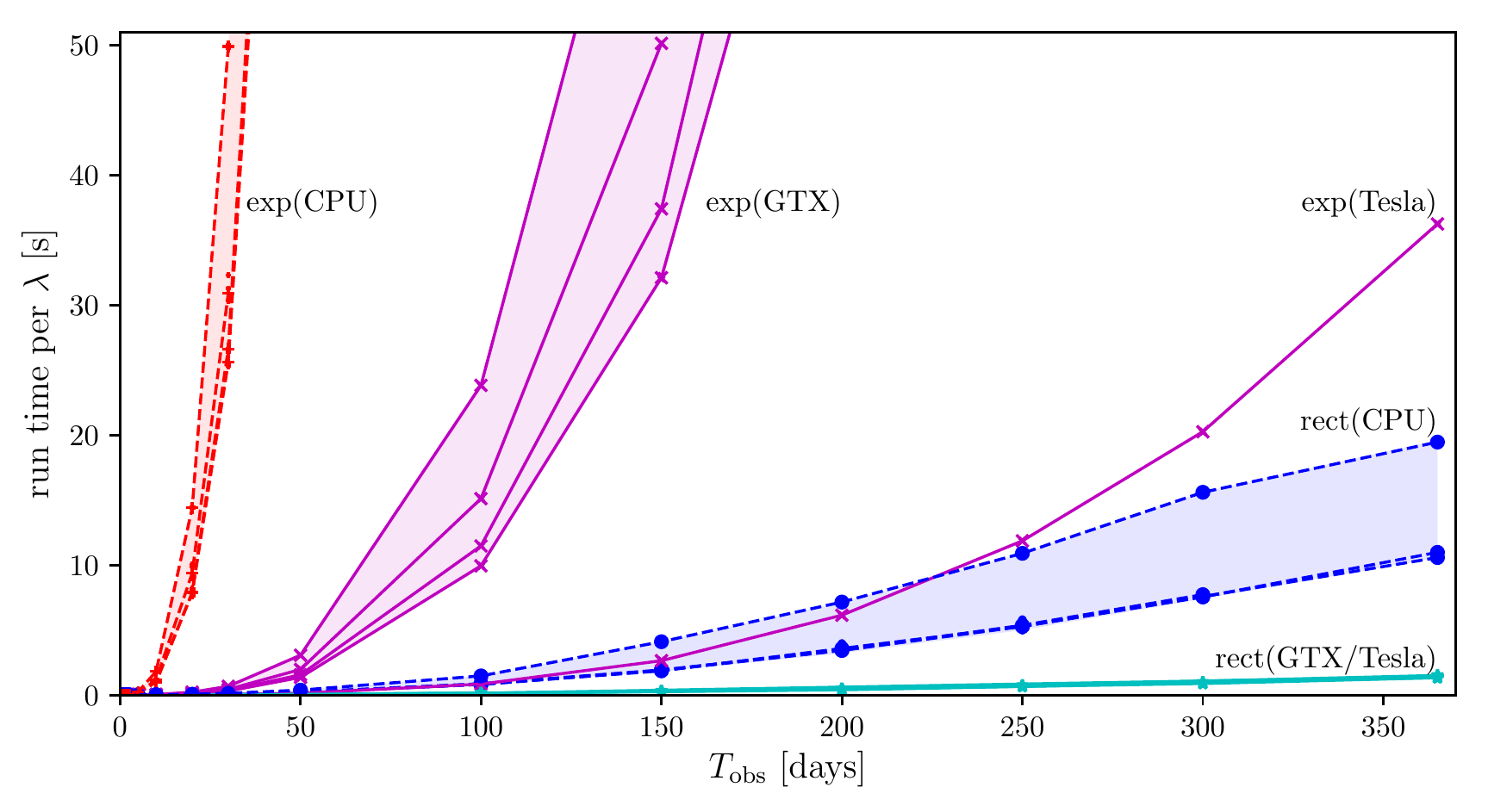}
  \caption{
   \label{fig:timing_all}
   Timing results for both rectangular and exponential transient windows,
   from CPU (\lalpulsar) and \gpu (\pycuda) implementations on various devices.
   The vertical axis gives the average run time
   per template $\Dop$.
   (Most test runs used 100 frequency bins,
   and a few used 1000 to check the consistency of averages.)
   Each solid/dashed line connects results
   from a specific implementation
   on a specific device,
   averaging over 3 or more runs at fixed $\Tobs$,
   and background shading indicates a specific window
   run on a family of related architectures.
   The exp(CPU) and rect(CPU) families collect results from
   the four different systems mentioned in Sec.~\ref{sec:timing},
   while exp(GTX) labels results from
   different Nvidia Geforce GXT 10x0 family devices,
   the single line labelled exp(Tesla)
   is from a Nvidia Tesla V100,
   and the rect(GTX/Tesla) results are plotted together
   as they are not significantly different.
  }
 \end{center}
\end{figure}

Since \gpu results for a single-template
(fixed $\Dop$)
analysis might be too pessimistic
because of startup overheads,
and in practice speedups are only relevant
for searches over broad $\Dop$ regions anyway,
we time searches over 100 frequency bins;
though for simplicity we assume a
fixed sky location and no spindown.
Timing results are summarised in \fref{fig:timing_all},
as average runtime per $\Dop$ template.
As an additional cross-check,
these results also include some runs at 1000 frequency bins,
which yield consistent timings per template.
Note that this is the total runtime of the search
(per template),
including the initial \lalpulsar computation
of the atoms which always runs on the CPU.

We find that
the \pycuda version provides speedups
of at least an order of magnitude
on \gpu[s] of the Geforce GTX 10x0 family
compared to the original \lalpulsar code on contemporaneous CPUs,
both for exponential and rectangular windows.
In the exponential case,
the Tesla V100 provides another similar jump in speed
over the GTX family,
bringing the cost of exponential-window transient searches
over hundreds of days
down to a similar cost
as with the standard rectangular \lalpulsar CPU implementation.

We can also more directly compare these measurements
to the timing model from Appendix A3 of~\citepgm.
We find that to cover arbitrary combinations of $\{\Tdata,\Nto,\Ntau\}$,
we need to somewhat generalise the model.
This is done in detail in our~\ref{sec:timing-model}.
However, for the timings presented in \fref{fig:timing_all},
we are in regimes where the cost for rectangular windows
is dominated by the $\Nto\,\Ntau$ scaling
and the cost for exponential windows
is dominated by the $\Nsums$ scaling.
The results,
converted to the `timing constants'
as introduced in~\citepgm,
are listed in table~\ref{tbl:timing_constants},
and are generally consistent with fits
of the more general timing model.

\begin{table}
 \caption{
  \label{tbl:timing_constants}
  Timing results from \fref{fig:timing_all},
  fit assuming dominant
  $\Nto\,\Ntau$ scaling for rectangular windows
  and $\Nsums$ scaling for exponential windows,
  and converted to the timing constants
  $\costrectPGM$, $\costexpPGM$
  as introduced by~\citepgm.
  Fit errors are $<1$\%.
  See \ref{sec:timing-model} for details
  and for complementary example fits
  of the more general timing model.
  \newline
  Note that \mbox{$\costexpPGM<\costrectPGM$} for the GPUs
  does not mean that the overall search
  for an exponential window
  is faster than for a rectangular window,
  as these constants are multiplied with different
  summation counters,
  see \eref{eq:cost-exp-PGM} and \eref{eq:cost-rect-PGM}.
 }
 \vspace{0.5\baselineskip}
 \begin{indented}
  \item[]
  \begin{tabular}{@{}llll}
\br
CPU/GPU    & $\costrectPGM$ [s]   & $\costexpPGM$ [s]   & \\
\mr
Core2Duo 2.6\,GHz & $4.2\cdot10^{-8}$ & $1.3\cdot10^{-7}$  & from~\citepgm \\
\hline
i5-6200U     & $6.4\cdot10^{-8}$ & $1.1\cdot10^{-7}$  & \\
Xeon X5675   & $3.5\cdot10^{-8}$ & $7.0\cdot10^{-8}$  & \\
Xeon E5-2630 & $3.4\cdot10^{-8}$ & $5.7\cdot10^{-8}$  & \\
Xeon E5-2650 & $3.6\cdot10^{-8}$ & $6.0\cdot10^{-8}$  & \\
\hline
GTX-1050     & $6.1\cdot10^{-9}$ & $1.4\cdot10^{-9}$  & \\
GTX-1060     & $4.8\cdot10^{-9}$ & $9.1\cdot10^{-10}$ & \\
GTX-1070     & $4.2\cdot10^{-9}$ & $7.3\cdot10^{-10}$ & \\
GTX-1080     & $4.4\cdot10^{-9}$ & $6.2\cdot10^{-10}$ & \\
Tesla-V100   & $4.3\cdot10^{-9}$ & $4.6\cdot10^{-11}$ & \\
\br
\end{tabular}
 \end{indented}
\end{table}

\subsection{Memory}
\label{sec:memory}

\gpu applications are often memory-limited.
However, for the transient \Fstat map,
we do not expect \gpu memory to be a significant constraint,
as we see in the following.
With the current approach,
the input atoms need to be transferred to \gpu memory
only for a single $\Dop$ parameter space point at a time,
then the $\Fmn(\Dop)$ matrix is computed and returned.
Hence, the peak \gpu memory usage of input plus output matrices is expected to be
\begin{equation}
 M [\mathrm{bytes}] = 4 \, ( \, 7\Nsft + \Nto \Ntau \, ) \,,
\end{equation}
where 4 bytes is the base size of a \sw{real32} number
in the underlying \numpy~\cite{numpy} package.
While the input array size grows only linearly with $\Nsft$,
assuming \mbox{$\resto=\restau=\Tsft$} the $\Fmn$ matrix grows quadratically
and will dominate memory usage at long $\Tobs$.
However, in practice one might want to choose an undersampling of $t_0$, $\tau$.

\begin{SCfigure}[50]
  \includegraphics[width=0.5\columnwidth]{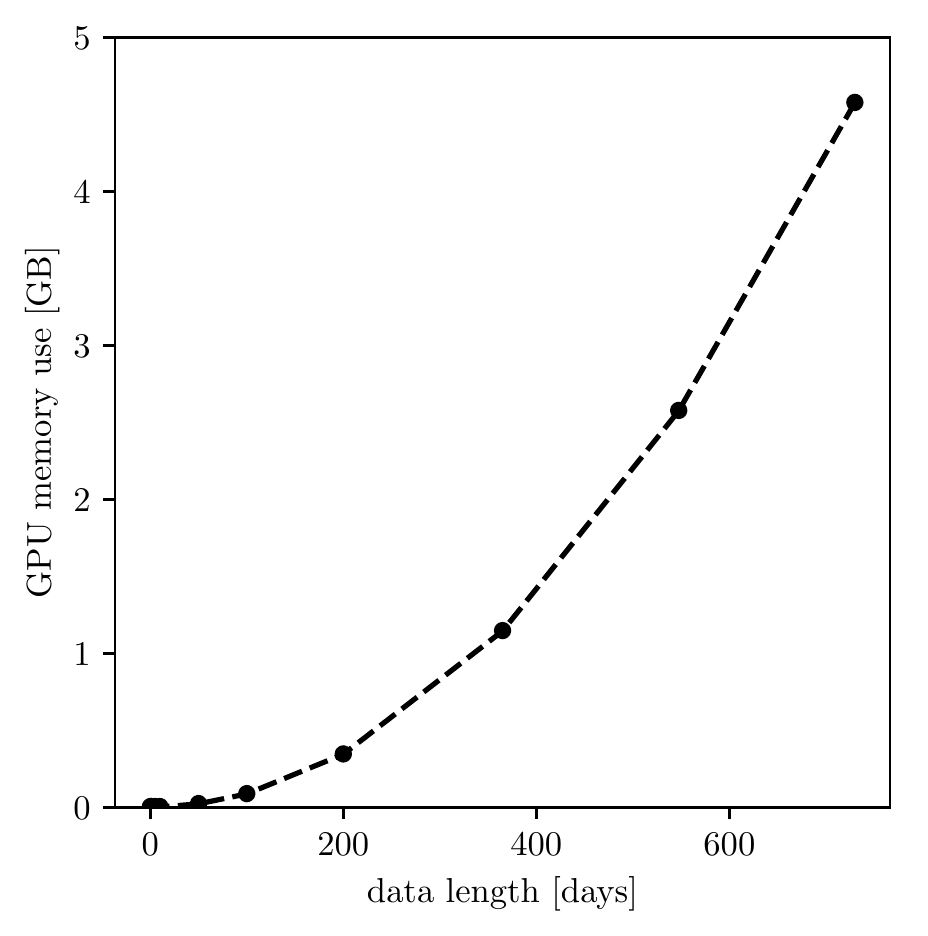}
  \caption{
   \label{fig:gpu_memory}
   \gpu memory usage
   on a GeForce GTX 1070 (8\,GB RAM)
   with CUDA V8.0.61.
   Measured with
   \mbox{$\Tsft=1800$\,s}
   and a resolution of
   \mbox{$\resto=\restau=\Tsft$}
   at all $\Tobs$.
   Each data point is the difference
   between the output of a call to
   \sw{pycuda.driver.mem\_get\_info()}
   right before
   allocating input and output arrays with \sw{pycuda.gpuarray},
   and a call right afterwards.
   The dashed line is the expected $4(7\Nsft + \Nto \Ntau )$ scaling (in bytes).
   As \mbox{$\Tobs\to0$},
   we find that the base memory use
   for the kernel itself
   (and any other possible overheads)
   seems to be only about 2--4\,MB.
  }
\end{SCfigure}

A comparison of this expectation with practical memory usage measurements
is presented in \fref{fig:gpu_memory}.
For \mbox{$\Tsft=1800$\,s} and \mbox{$\resto=\restau=\Tsft$},
the memory usage reaches only about 1.1\,GB for a year of data,
and with undersampling even much longer data sets
would remain easily feasible on current \gpu[s],
even when multiple jobs need to run on a single device.

\subsection{Accuracy}
\label{sec:accuracy}

The original \lalpulsar implementation
is already using single precision
for the atoms and the \Fstat itself,
so in contrast to some other \gpu use cases~\cite{Navarro:2014gpu}
it was not necessary to reduce the code's internal precision
for the \pycuda version.
However, the \Fstat algorithm is already known
to produce slightly different numerical results
on different CPU platforms,
so it is worth checking the typical amount of differences in
the transient \Fstat between \lalpulsar and \pycuda versions.

\begin{figure}
 \begin{center}
  \includegraphics[width=\columnwidth]{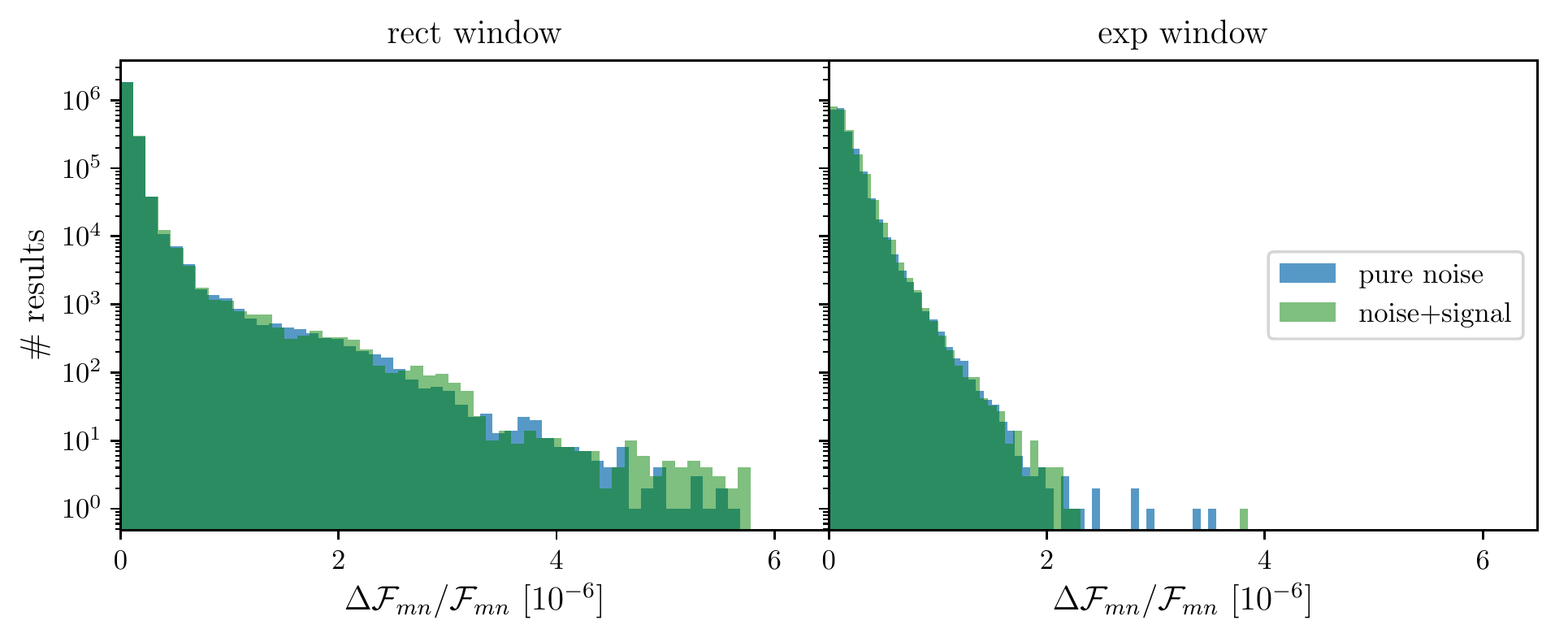}
  \caption{
   \label{fig:accuracy_hists}
   Comparison of $\Fmn$ computed
   with the \lalpulsar and \pycuda implementations.
   Each histogram gives the differences $\Delta\Fmn$
   between the two implementations
   for a certain transient window,
   and either for pure Gaussian noise
   or also including a
   (confidently detectable,
   \mbox{$\max2\F\approx263$})
   signal injection
   with matching window function.
   The histograms are taken over all individual $\Fmn$ values
   for 1000 frequency bins
   over a 1-day data set with
   \mbox{$\Tsft=1800$}.
   The \gpu for this test was a GeForce GTX 1070.
  }
 \end{center}
\end{figure}

As demonstrated for a particular test case in \fref{fig:accuracy_hists},
we typically find negligibly small differences,
not larger than other implementation- and platform-dependent variations
in the \Fstat known from other work (e.g.~\cite{Prix:Fstat-bias}).

One implementation detail to note is that
in the exponential case
the \lalpulsar implementation
uses a lookup-table (LUT) based `fast exponential' function.\footnote{As
of the writing of this paper,
with git tag d0d28012640f649bd910367c027385556689ed38
of the \url{https://git.ligo.org/lscsoft/lalsuite/} repository.}
This can actually lead to differences with \pycuda
of up to $\sim10$\%,
but \fref{fig:accuracy_hists} shows results after replacing it
with the \sw{exp()} function of the C standard library,
thus verifying that the difference
did not come from a loss of accuracy with the new \pycuda implementation.

\section{Conclusion and applications}
\label{sec:concl}

The significant speedup
achieved with our \pycuda implementation
of the transient \Fstat
will allow for a wider scope of searches
for long-duration transient \gw[s].
We now discuss a few example applications
that would be hard resource-wise,
or even prohibitive,
on CPUs
but could become viable with GPUs.

Let us first consider the natural use case
of a \gw data analysis triggered by radio observations
of a pulsar glitch.
Quasi-monochromatic \gw emission,
which the \Fstat is sensitive to,
could be associated with the post-glitch relaxation.
Depending on the pulsar, this can have timescales
of days to months~\cite{Lyne:2000sta,Haskell:2013goa}.
As a simple transient search setup,
assume we look at a single fixed $t_0$ and at
\mbox{$\tau\in[2\Tsft,\,\Tobs=4\,\mathrm{months}]$}
with \mbox{$\delta\tau=\Tsft=1800$s}.
With these parameters,
we find a Tesla-V100 GPU outperforms a Xeon-E5 CPU
by a runtime factor of $\approx10$ for rectangular windows
and $\approx2300$ for exponential windows.
Still, with a single \gw search template
matching the post-glitch radio timing solution
(at \mbox{$\fgw=2\fspin$}),
such an analysis would be computationally trivial
even on a single CPU
($\approx1700\mathrm{s}$ for exponential windows).

However, it would be reasonable to allow for some
mismatch between the radio timing and \gw frequency evolution
due to the perturbed state of the \ns after a glitch.
For comparison, the `narrow-band' search
for CWs from known pulsars in the first aLIGO run~\cite{Abbott:2017cvf}
(using 121 days of data) covered some ranges in
frequency $\freq$ and spindown $\fdot$
for each of its 11 targets,
with totals of e.g.
$2.2\cdot10^{6}$
templates 
for the Vela pulsar
and
$1.6\cdot10^{8}$
for the crab pulsar.
(Using the corrected Table I
as per the erratum~\cite{Abbott:2018err}
to~\cite{Abbott:2017cvf}.)

Multiplying these numbers of templates with
the per-template transient \Fstat cost
(which in this setup again dominates
over the rest of the \Fstat search code),
we find that a single Tesla-V100 GPU
could perform an exponential-window transient analysis
over the Vela band in less than 3 weeks,
while the same analysis would take 120 years
on a single Xeon-E5-like CPU;
or equivalently would require over 2300 CPUs
to only take the same 3 weeks as the single GPU.

Meanwhile, for the wider Crab analysis range
(which is due to its strong spindown),
even the GPU would still need 4 years
(compared to 9000 years for a single CPU).
While we can trivially further parallelise the problem
by splitting the $(\freq,\fdot)$ space over multiple GPUs,
only a small number of such devices are available on current computing clusters.
We can gain some more speed-up by reducing the sampling in $\tau$,
but in the end the parameter space for the Crab
would need to be somewhat reduced in practice.

In summary,
performing routine transient \Fstat analyses
of \emph{all} observed glitches
in known galactic pulsars
during a \gw observation run --
with reasonably wide $\freq$ and $\fdot$ bands
(similar to those used in~\cite{Abbott:2017cvf},
or only slightly reduced) --
becomes feasible with a few dedicated GPU systems.

Similar estimates apply
when considering the follow-up~\cite{Shaltev:2013kqa,Papa:2016cwb,Ashton:2018ure}
of significant or marginal detection candidates
produced by wide-parameter space \cw searches~\cite{Riles:2017evm}.
Even though those searches target perfectly persistent signals,
they can also produce candidates
if there are sufficiently strong transient events in the data~\cite{Keitel:2015ova}.
A comprehensive transient-aware follow-up,
with the goal of either verifying
the presumed persistent nature
or uncovering a transient signal instead,
needs to not only target
the exact phase-evolution parameters $\Dop$ of the candidate,
but search a wider band around it
to account for degeneracies
with the transient evolution parameters.
Reducing the computational cost of each candidate's follow-up
directly translates into a larger number of candidates that
can be analysed,
so that the overall threshold of the \cw search can be lowered
and a better search sensitivity can be achieved.

The data length and $\{t_0,\tau\}$ ranges in this scenario
can be longer than in the EM-triggered post-glitch scenario:
the aLIGO runs O1--O3 took data for / are scheduled for
4, 9 and 12 months respectively~\cite{Aasi:2013wya},
and for the follow-up of a strong candidate
data from multiple observing runs could get combined.
The range of phase evolution parameters $\Dop$
that should be searched for full coverage
depends on the exact setup of the \cw search
and on possible intermediate follow-up steps;
but the scaling of the transient \Fstat cost
is similar as in the EM-triggered case
(see \ref{sec:timing-model} for the full timing model)
and the improvements in accessible search volume
using a small number of GPUs over CPUs will be similar.

In the longer term,
untriggered all-sky searches
for long-duration transients
are of high interest.
Similarly to all-sky \cw searches,
they have the potential
to discover a population of electromagnetically dark
\ns[s],
for example glitching pulsars
with their beam pointed away from Earth.
The sensitivity of all-sky searches is directly
limited by how densely they can cover the $\Dop$
parameter space at a fixed computational budget.~\cite{Prix:2012yu,Wette:2011eu}.
Hence, adding transient parameters at first significantly
reduces the overall sensitivity of a blind search.
But speeding up the transient part
by orders of magnitude
could still make a combined search for \cw[s] and transients feasible
in the long run,
when large numbers of GPUs become available
in high-performance clusters
or through Einstein@Home~\cite{EatH}.
In practice, though,
the more promising approach for blind transient searches
might be to apply
a cheap add-on transient modification,
like that introduced in~\cite{Keitel:2015ova},
to a semi-coherent \cw algorithm
as a first search stage,
then apply the fully-coherent transient \Fstat
only in a follow-up step.

In any of these scenarios,
while we have focussed on the fact that
the \pycuda version can bring down the cost
of exponential-windowed transients
significantly,
the cost for rectangular windows always remains smaller,
so that in practice whenever exponential windows are feasible,
it is also cheap and natural
to run \emph{both} analyses
and evaluate a posteriori which one fits the data better.
Different window functions for the amplitude evolution
could also be considered,
and would generically follow
the GPU kernel grid setup
and timing model
for the exponential window,
since it does not assume any function-specific optimisations.

\ack{The authors would like to thank
Reinhard Prix for feedback on the manuscript
as well as him and Karl Wette
for development and continued support
of the underlying LALSuite \Fstat code.
Thanks to Chris Messenger and Stuart Anderson
for support with the GPU test systems
at Glasgow and CIT.
\\
DK was funded under the EU Horizon2020 framework
through the Marie Sk\l{}odowska-Curie grant agreement 704094 GRANITE,
and would also like to acknowledge the
Horizon2020 ASTERICS-OBELICS International School (Annecy 2017)
that inspired this investigation into \pycuda and \gpu applications.
}

\appendix
\section*{Appendix}

\section{Generalising the PGM2011 timing model}
\label{sec:timing-model}

Here we revisit the timing model
for computing $\Fmn$ maps
introduced in Appendix A3 of~\citepgm.
Their equations (A13) and (A14) give the
computing cost for a single-$\Dop$ $\Fmn$ map
with either exponential
\begin{equation}
  \label{eq:cost-exp-PGM}
  \hspace{-2.2cm}
  \costFmapexp \approx \costexpPGM \, \frac{\Delta t_0}{\resto} \, \frac{\Delta\tau}{\restau} \, \frac{(\taumin + \Delta\tau/2)}{\Tsft}
               \approx \costexpPGM \, \Nto                      \, \Ntau                      \, \frac{(\taumin + \Delta\tau/2)}{\Tsft}
\end{equation}
or rectangular window functions:
\begin{equation}
  \label{eq:cost-rect-PGM}
  \hspace{-2.2cm}
  \costFmaprect =       \costrectPGM \, \frac{\Delta t_0}{\resto} \, \frac{(\tau_{\min} + \Delta\tau)}{\Tsft}
                \approx \costrectPGM \, \Nto                      \, \frac{(\tau_{\min} + \Delta\tau)}{\Tsft}
                =       \costrectPGM \, \Nto                      \, \frac{\tau_{\max}}{\Tsft}  \,.
\end{equation}
The timing constants $\costexpPGM$ and $\costrectPGM$
are interpreted as the cost to compute
the (weighted) sums over atoms at each step.
The exponential model corresponds to a `generic' case
where all quantities have to be re-evaluated at each step,
while the rectangular case reuses partial sums as discussed before.

We note now that this formulation of the timing model
does not explicitly include the cost of computing
the antenna pattern matrix determinant $\Dd$
and the $\F$-statistic itself,
which is done once for each $(m,n)$ pair
after all sums have been computed
and hence is independent of the window function choice.\footnote{In its scalings,
this extra cost
is degenerate with
the marginalisation cost $\costmarg$ of PGM's Eq. (15),
in search code executions were both $\Fmn$ maps and marginal Bayes factors are computed;
so it was effectively included in PGM's overall code timing,
but just attributed to a different part of the model.}
We can include this contribution
by adding a term $+\costF\,\Nto\Ntau$ to both cases.
It will be very subdominant for exponential windows,
where the summations term grows much faster than $\Nto\Ntau$,
but can be relevant for rectangular windows
where the summations term is more efficient.

Another small contribution to the timing model is from setup
and index-lookup costs that scale with the total number of SFTs
handed to the \Fstat-map function;
for completeness we include a common term $\costSFTs\,\Nsft$.

In addition,
Eqs.~\eref{eq:cost-exp-PGM} and~\eref{eq:cost-rect-PGM}
only hold true
if the full range of transient signal durations
explored by the $\Fmn$ map
is fully contained within the available data range,
that is when
\mbox{$\tomax+\taumax < T_0 + \Tdata$}.
(We call this the `embedded' case below.)
Otherwise, i.e. if some of the transient windows
overlap the end of the available data,
by convention the \lalpulsar code still returns
results for the full rectangular $\Fmn$ matrix,
but truncates the atoms summations.
Thus, the total computing cost in such cases
is lower than estimated by
Eqs. \eref{eq:cost-exp-PGM}, \eref{eq:cost-rect-PGM}
and using them to fit the timing constants from
runtime measurements as in Sec.~\ref{sec:timing}
would yield inconsistent results.

Hence, we generalise the PGM timing model
by introducing $\Nsums$ as the effective number of summation steps
for an $\Fmn$ map,
which depends on the window type,
$\Tdata$, and the ranges of both $t_0$ and $\tau$:
\begin{equation}
  \label{eq:cost-exp-Nsums}
  \costFmapexp  \approx \costSFTs\,\Nsft + \costexpsums  \, \Nsumsexp  + \costF \, \Nto \, \Ntau \,,
\end{equation}
\begin{equation}
  \label{eq:cost-rect-Nsums}
  \costFmaprect \approx \costSFTs\,\Nsft + \costrectsums \, \Nsumsrect + \costF \, \Nto \, \Ntau \,.
\end{equation}
For rectangular windows, we have
\begin{equation}
 \label{eq:Nsums-rect}
 \Nsumsrect = \sum\limits_{m=1}^{\Nto} \frac{\min(\Tdata-t_{0\,m},\,\taumax)}{\Tsft} \,,
\end{equation}
which reduces to PGM's
\mbox{$\Nsumsrect = \Nto \taumax / \Tsft$} 
in the special `embedded' case
that PGM considered,
and to
\mbox{$\Nsumsrect = 0.5 \Nto \taumax / \Tsft$}
in the special case of
\mbox{$\Nto\resto=\Ntau\restau=\Tdata-2\Tsft$}
that we used for the timing results in Sec.~\ref{sec:timing}.

For exponential windows,
we also need to note that the current code's convention,
as introduced in Eq. (18) of~\citepgm,
is that an exponential window with duration parameter $\tau$
covers an effective length of $3\tau/\Tsft$ atoms.
(The exponential decay is not cut off after only one,
but after three e-folds,
where the remaining \snr would be much more negligible.)
Hence,
PGM's original timing constant
$\costexpPGM$ effectively contains a factor of 3
(from counting all steps in $\tau$)
that we now include in $\Nsumsexp$ instead:
\begin{equation}
 \label{eq:Nsums-exp}
 \Nsumsexp = \sum\limits_{m=1}^{\Nto} \sum\limits_{n=1}^{\Ntau} \frac{\min(\Tdata-t_{0\,m},\,3\,\tau_n)}{\Tsft} \,.
\end{equation}
In the `embedded' case this reduces to
\mbox{$3\Nto \sum\limits_{n=1}^{\Ntau} \tau_n/\Tsft = 3\Nto\Ntau(\taumin+0.5\Delta\tau)/\Tsft$},
equivalent to PGM's result up to the factor of 3.

In practice,
on each architecture
we can use these more general
equations~\eref{eq:cost-exp-Nsums}--\eref{eq:Nsums-exp}
to fit the four timing constants
$\{\costSFTs,\costF,\costrectsums,\costexpsums\}$
from a variety of setups
(in terms of $\Tdata$, $[\tomin,\tomax]$, $[\taumin,\taumax]$),
then consider the special `embedded' case\footnote{The example
used for timing in Appendix A3 of~\citepgm
is 1 year of data
with $\tau\in[0.5,14.5]\,$days;
which means that with t0 close to the end of the year
and $\taumax=14.5\,$days the overlap is at most 4\%
and the deviations from the fully-embedded special case
used for this comparison are smaller than typical timing uncertainties.}
(and \mbox{$\Ntau \gg 1$}, \mbox{$\taumax \gg \taumin$})
to directly compare to~\citepgm
by
\begin{equation}
  \label{eq:cost-rect-convert}
  \hspace{-2cm}
 \costrectPGM =        \costrectsums + \frac{\Tsft}{\taumax} \left( \costSFTs\frac{\Nsft}{\Nto} + \costF\,\Ntau \right)
              \approx  \costrectsums + \costF \,,
\end{equation}
\begin{equation}
  \label{eq:cost-exp-convert}
  \hspace{-2cm}
 \costexpPGM  =       3\costexpsums  + \frac{\Tsft}{\taumin+0.5\Delta\tau} \left( \costSFTs \frac{\Nsft}{\Nto\Ntau} + \costF \right)
              \approx 3\costexpsums  + \frac{2}{\Ntau} \costF
              \approx 3\costexpsums \,.
\end{equation}

Using a set of timing runs that in addition to those in section~\ref{sec:timing}
also cover many different combinations of $\{\Tdata,\Nto,\Ntau\}$,
and also measuring \textit{only} the executation time of the actual \Fstat map function
(while in section~\ref{sec:timing} the whole search call is timed,
including the contribution of computing the atoms,
which is usually subdominant but not in the limit of low $\Nto\,\Ntau$ and $\Nsums$),
we do a detailed fit of the full timing model of \eref{eq:cost-exp-Nsums} and \eref{eq:cost-rect-Nsums},
in the following iterative steps to ensure convergence:
\begin{enumerate}
 \item fit the $\costSFTs\,\Nsft$ term only to short data sets with $\Nsums \leq 100$
 \item using this fixed $\costSFTs$, fit $\costrectsums\,\Nsumsrect + \costF\,\Nto\,\Ntau$ for rectangular windows
 \item using fixed $\costSFTs$ and $\costF$, fit $\costexpsums\,\Nsumsexp$ for exponential windows
\end{enumerate}
We find e.g.
\begin{equation}
 \hspace{-2.5cm}
 \costFmaprect \approx \left( (2.80\pm0.03)\Nsft + (0.96\pm0.08)\Nsumsrect + (5.59\pm0.06)\Nto\Ntau \right) 10^{-8}\,\mathrm{s}
\end{equation}
\begin{equation}
 \hspace{-2.5cm}
 \costFmapexp  \approx \left( (2.80\pm0.03)\Nsft + (3.55\pm0.03)\Nsumsexp  + (5.59\pm0.06)\Nto\Ntau \right) 10^{-8}\,\mathrm{s}
\end{equation}
for the i5-6200U laptop CPU
(corresponding to PGM constants
\mbox{$\costrectPGM=(6.36\pm0.08)10^{-8}$}
and \mbox{$\costexpPGM=(1.07\pm0.01)10^{-7}$});
and
\begin{equation}
 \hspace{-2.5cm}
 \costFmaprect \approx \left( (2.59\pm0.02)\Nsft + (0.27\pm0.02)\Nsumsrect + (3.09\pm0.02)\Nto\Ntau \right) 10^{-8}\,\mathrm{s}
\end{equation}
\begin{equation}
 \hspace{-2.5cm}
 \costFmapexp  \approx \left( (2.59\pm0.02)\Nsft + (2.22\pm0.03)\Nsumsexp  + (3.09\pm0.02)\Nto\Ntau \right) 10^{-8}\,\mathrm{s}
\end{equation}
for the Xeon X5675 workstation CPU
(corresponding to PGM constants
\mbox{$\costrectPGM=(3.26\pm0.02)10^{-8}$}
and \mbox{$\costexpPGM=(6.67\pm0.08)10^{-8}$}).
These results agree reasonably well
with those obtained on the same systems,
but with fixed $\Nto,\Ntau$ in relation to $\Tdata$
and with simplified fits,
as presented in table~\ref{tbl:timing_constants}.
While the error bars from fitting alone
appear too small to explain the remaining differences of 0.5--7\%,
it is likely that variations in system configuration and load
between timing runs are the main culprit.

{
 \phantomsection
 \addcontentsline{toc}{chapter}{References}
 \small
 \bibliography{biblio_tCWgpu.bib}
}

\end{document}